\documentclass[twocolumn, amssymb,nobibnotes,aps,prl,nopacs,superscriptaddress]{revtex4-1}

\usepackage{graphicx}
\usepackage[table]{xcolor}
\usepackage{booktabs}
\usepackage{epstopdf}
\usepackage{verbatim}

\begin{document}

\title{Temporal structure of attosecond pulses from laser-driven coherent synchrotron emission}

\author{S.\ Cousens\footnote{Email: s.cousens@qub.ac.uk}}
\affiliation{Centre for Plasma Physics, Department of Physics and Astronomy, Queen's University Belfast, BT7 1NN, United Kingdom}

\author{B.\ Reville}
\affiliation{Centre for Plasma Physics, Department of Physics and Astronomy, Queen's University Belfast, BT7 1NN, United Kingdom}

\author{B.\ Dromey}
\affiliation{Centre for Plasma Physics, Department of Physics and Astronomy, Queen's University Belfast, BT7 1NN, United Kingdom}

\author{M.\ Zepf}
\affiliation{Centre for Plasma Physics, Department of Physics and Astronomy, Queen's University Belfast, BT7 1NN, United Kingdom}
\affiliation{Helmholtz Institut Jena, Fr\"{o}belstieg 3, 07743 Jena, Germany}

\date{\today}
\begin{abstract}
The microscopic dynamics of laser-driven coherent synchrotron emission transmitted through thin foils are investigated using particle-in-cell simulations.  For normal incidence interactions, we identify the formation of two distinct electron nanobunches from which emission takes place each half-cycle of the driving laser pulse. These emissions are separated temporally by 130 attoseconds and are dominant in different frequency ranges, which is a direct consequence of the distinct characteristics of each electron nanobunch.  This may be exploited through spectral filtering to isolate these emissions, generating electromagnetic pulses of duration $\sim70~\mathrm{as}$. \end{abstract}
\maketitle


The generation of attosecond pulses of extreme-ultraviolet (XUV) radiation via laser-matter interactions has been demonstrated as a powerful tool for the observation of ultrafast atomic and electronic phenomena \cite{corkum2007, krausz2009}. Experimental and numerical studies have shown that emissions of particularly high XUV photon flux are produced by focusing a laser pulse to relativistic intensities ($I\lambda_L^2 > 1.37\times10^{18}~\mathrm{W~cm^{-2}~\mu m^2}$) onto either a thick solid target \cite{bulanov1994,gibbon1996,lichters1996, PhysRevLett.99.085001, PhysRevE.74.046404, PhysRevLett.99.085001} or a thin nanofoil \cite{Hassner:97, PhysRevLett.92.185001, PhysRevE.72.036413, 1367-2630-11-11-113028, dromey2012coherent,1367-2630-15-1-015025,11ca994d232642f692b19e1fd1779f64, PhysRevLett.100.125005}.  In either case, a train of attosecond bursts is emitted by nanometer-scale bunches of electrons (nanobunches) which are driven periodically by the laser pulse into curved, synchrotron-like trajectories at the front surface of the resulting plasma \cite{dromey2012coherent,1367-2630-15-1-015025,:/content/aip/journal/pop/17/3/10.1063/1.3353050,mikhailova2012}.

However, application of this radiation to time-resolved measurements of ultrafast phenomena requires an isolated pulse, otherwise there will be an uncertainty as to which pulse from the train is probing the process. Polarization gating techniques applied to laser-solid interactions \cite{1367-2630-10-2-025025, PhysRevLett.112.123902,yeunggatingtobepublished} have shown recent promise in reducing the number of cycles during which emission occurs, while spatially selecting the pulses \cite{wheeler2012} may provide an alternative or a complement to these methods. If a pulse has been successfully isolated then the potential temporal resolution of this source is only limited by the structure of that pulse.  However, for oblique laser incidence, simulation results have shown that the reflected radiation may contain an unavoidable sub-structure which limits the temporal resolution of attosecond pulses produced in this geometry \cite{multibunching2011}.

In this Letter we provide the first in-depth numerical study of the temporal structure of coherent synchrotron emission transmitted through nanofoils irradiated at normal incidence. Our particle-in-cell simulation results show that two distinct electron nanobunches contribute to the emission, producing a double-pulse structure separated by $130~\mathrm{as}$. However, we demonstrate that in this geometry this sub-structure does not necessarily place a strict lower limit on the achievable pulse duration. Our simulation results show that the characteristics of the emitting nanobunches are different, and emit radiation with distinct spectral shapes.  We show that the relative intensity of the sub-pulses may be controlled by appropriate spectral filtering, resulting in a single pulse of duration $\sim70~\mathrm{as}$, capable of temporally resolving a wide variety of ultrafast phenomena.


To investigate coherent synchrotron emission (CSE) from relativistic laser-nanofoil interactions, a simulation was performed using the 1D particle-in-cell code PICWIG \cite{1367-2630-10-2-025025}.   The simulation models a $5 ~\mathrm{fs}$ FWHM in intensity, linearly polarized laser pulse, with a central wavelength of $\lambda_L = 800~\mathrm{nm}$ and $a_0 = 12$.  This pulse is normally incident onto an overdense, $n_e = 100~\mathrm{n_c}$, fully ionized carbon plasma, $160~\mathrm{nm}$ thick, with an additional $80~\mathrm{nm}$ linear density ramp at the front of the target.  Here $\mathrm{n_c} = \omega_L^2m_e\epsilon_0/e^2$ is the critical density, with $\omega_L$ being the laser frequency.   The $20\mathrm{\lambda_L}$ simulation box consists of 1000 cells/$\mathrm{\lambda_L}$, and initially 300 macro-electrons per cell.  The ions are assumed to be immobile.

\begin{figure}
\includegraphics[scale = 0.85]{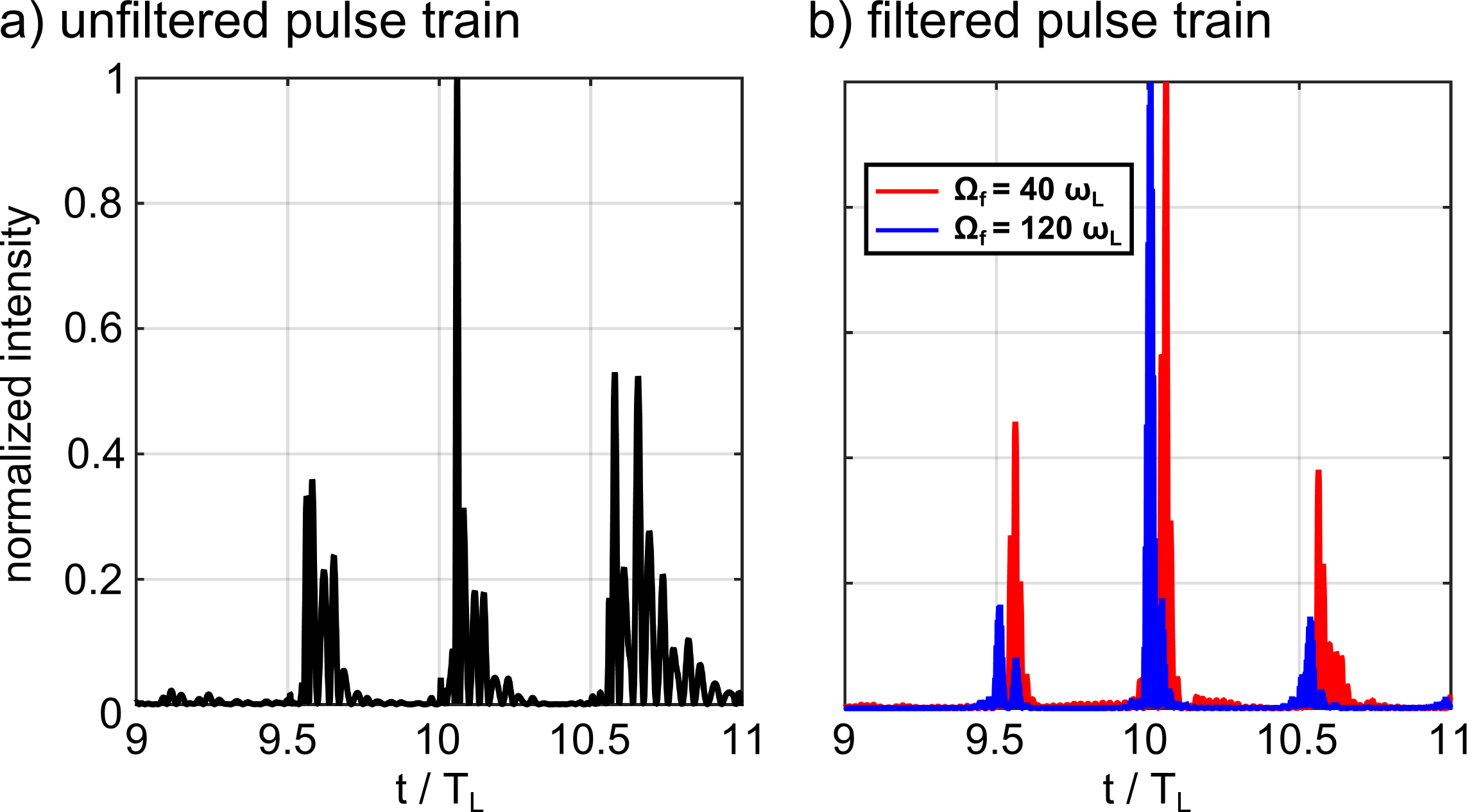}\\
\includegraphics[scale = 0.85]{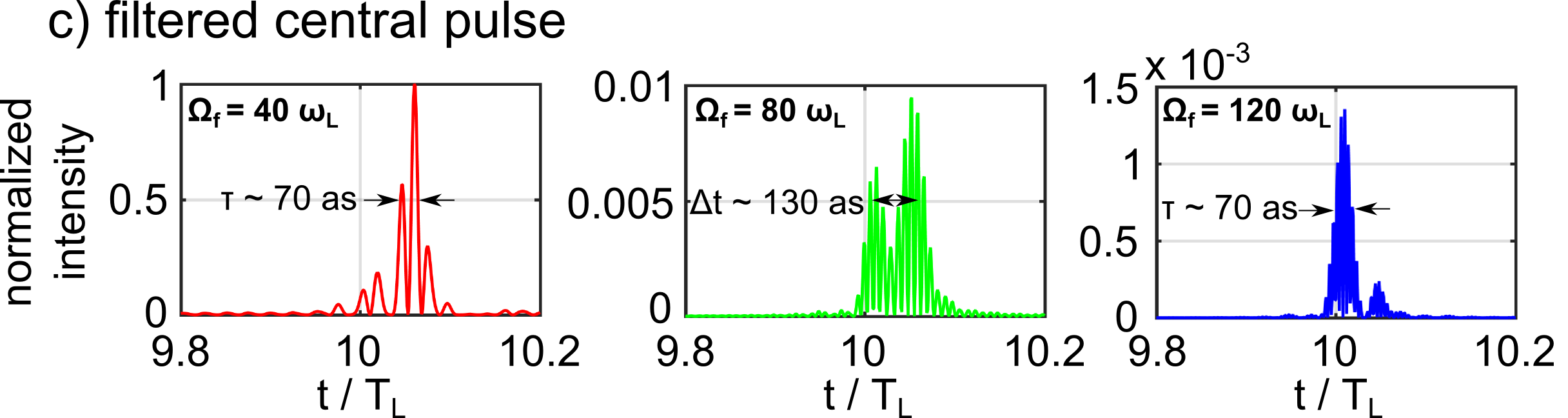}\\
\caption{Part a) shows the pulse train collected at the rear side of the target during a simulated relativistic laser-nanofoil interaction (simulation parameters given in text). Part b) shows this train, viewed through two different frequency windows of width $\Delta\omega = 40~\omega_L$, centered at $\Omega_f = 40~\omega_L$ and $\Omega_f = 120~\omega_L$. Part c) shows a closer view of the central pulse from this train for these spectral ranges and an intermediate range centered on  $\Omega_f = 80~\omega_L$. These results suggest that each individual burst of CSE is composed of two distinct attosecond sub-pulses, which dominate at different frequency ranges, and are separated temporally by $\Delta t\approx130~\mathrm{as}$.\label{fig1}}
\end{figure}

The train of pulses collected at the rear side of the target during this simulation is shown in Fig.~\ref{fig1}a). Here $T_L = 2\pi/\omega_L$ is the laser period. From this figure, it is clear that the pulses possess a twice-per-cycle periodicity.  Common to each of these pulses is the underlying structure of coherent synchrotron emission, which we investigate here. Firstly, we view this train through various frequency windows.  To do this, we apply a rectangular window function of width $\Delta\omega = 40~\omega_L$, centered at frequency $\Omega_f$, to the spectrum of the pulse train, with all other frequencies outside this window excluded. Fig.~\ref{fig1}b) shows the pulse train for the filter positioned at both $\Omega_f/\omega_L = 40$ (red) and $\Omega_f/\omega_L = 120$ (blue). It is clear that the attosecond pulses in these two frequency windows do not overlap in time, with the higher frequency range preceding the lower.  Fig.~\ref{fig1}c) provides a closer view of the central pulse in the train for these two spectral ranges, from which it is clear that two distinct sub-pulses are present, each of duration $\tau \approx 70~\mathrm{as}$ which are temporally offset by $\Delta t \approx 130~\mathrm{as}$.   Also shown is the pulse corresponding to an intermediate range, $\Omega_f/\omega_L = 80$ (green), for which the contributions of these two sub-pulses are comparable and a longer double-pulse structure emerges.

\begin{figure}
\includegraphics[scale = 0.85]{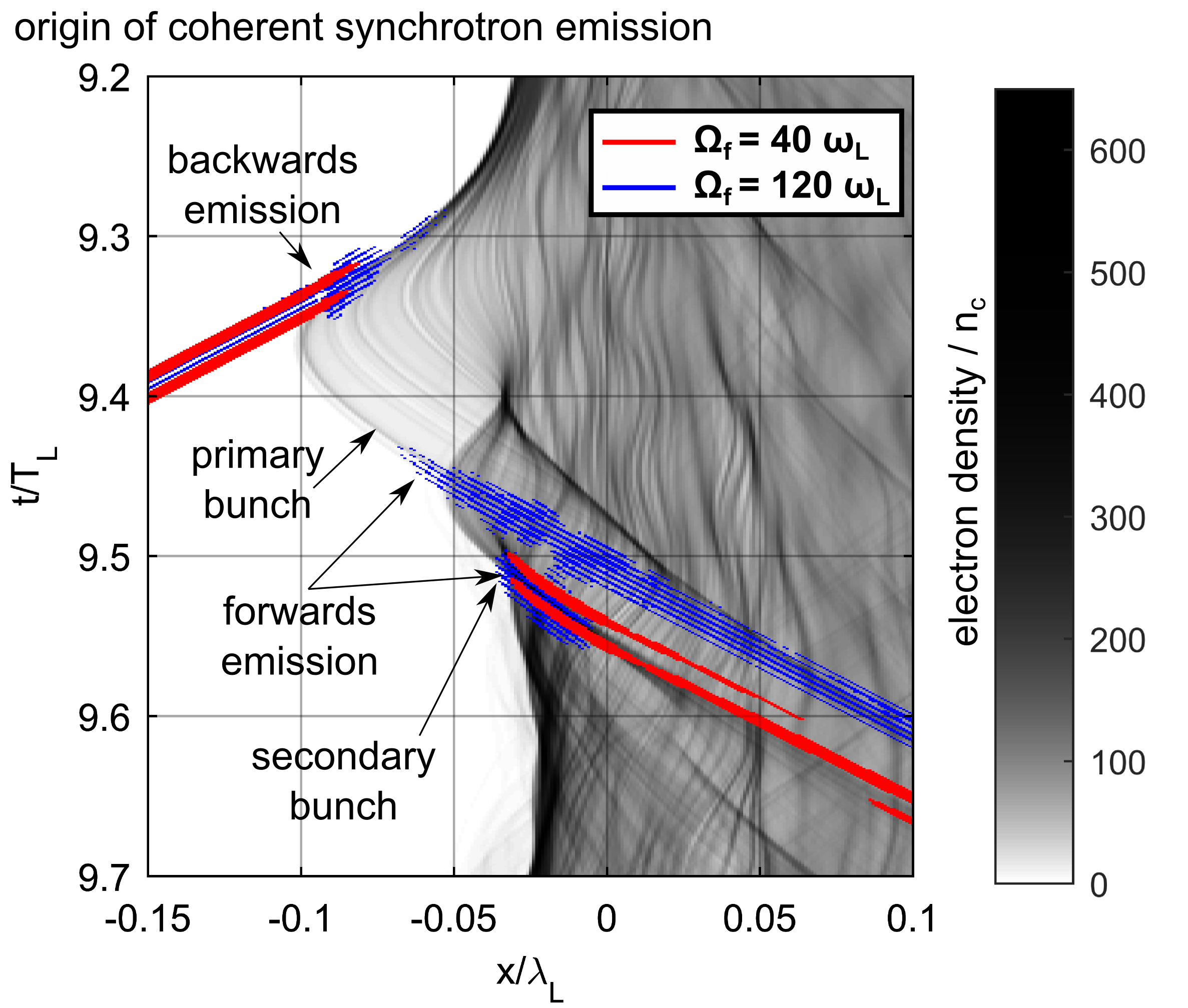}
\caption{The electron density (gray-black), overlaid with the electric field intensity for the $\Delta\omega = 40~\omega_L$ filter window centered at $\Omega_f = 40~\omega_L$ (red) and  $\Omega_f = 120~\omega_L$ (blue).  In order to highlight the origin of the strongest radiation, only the most intense $60\%$ of radiation is shown, in block color. By tracing these pulses to their source it is seen that, in contrast to the reflected emission, the forward pulses are generated by two distinct electron bunches. \label{fig2alt}}
\end{figure}

To trace the origins of these pulses, the evolution of the electron density at the front surface of the plasma, for half a laser period surrounding the emission of this central pulse, is shown in Fig.~\ref{fig2alt} (gray-black).  Overlaid on this is the electric field intensity for the two spectral ranges, centered at $\Omega_f/\omega_L = 40$  (red) and $\Omega_f/\omega_L = 120$ (blue).  Similar to a conventional synchrotron, the emissions occur when the electrons are traveling towards the observer at relativistic velocities. For the reflected radiation, both spectral ranges are shown to originate from the same point, as the electrons are traveling out from the plasma in the $-{\bf \hat{x}}$ direction.  In the transmitted direction however, it is clear that there are two distinct bunches of electrons that are acting as sources of radiation.  The first or \emph{primary} electron bunch is seen to be the dominant source of radiation for higher frequencies, whereas the \emph{secondary} electron bunch dominates production of the lower frequencies. We note that the formation of this secondary bunch is clearly visible from simulation results in a number of previous publications \cite{lichters1996, 1367-2630-10-2-025025, PhysRevLett.112.123902, 1367-2630-11-11-113028, 1367-2630-15-1-015025}.  However, the mechanism behind its origin, as well a comparison between its emission and that from the primary bunch has yet to be presented.  

\begin{figure}
\includegraphics[scale = 0.75]{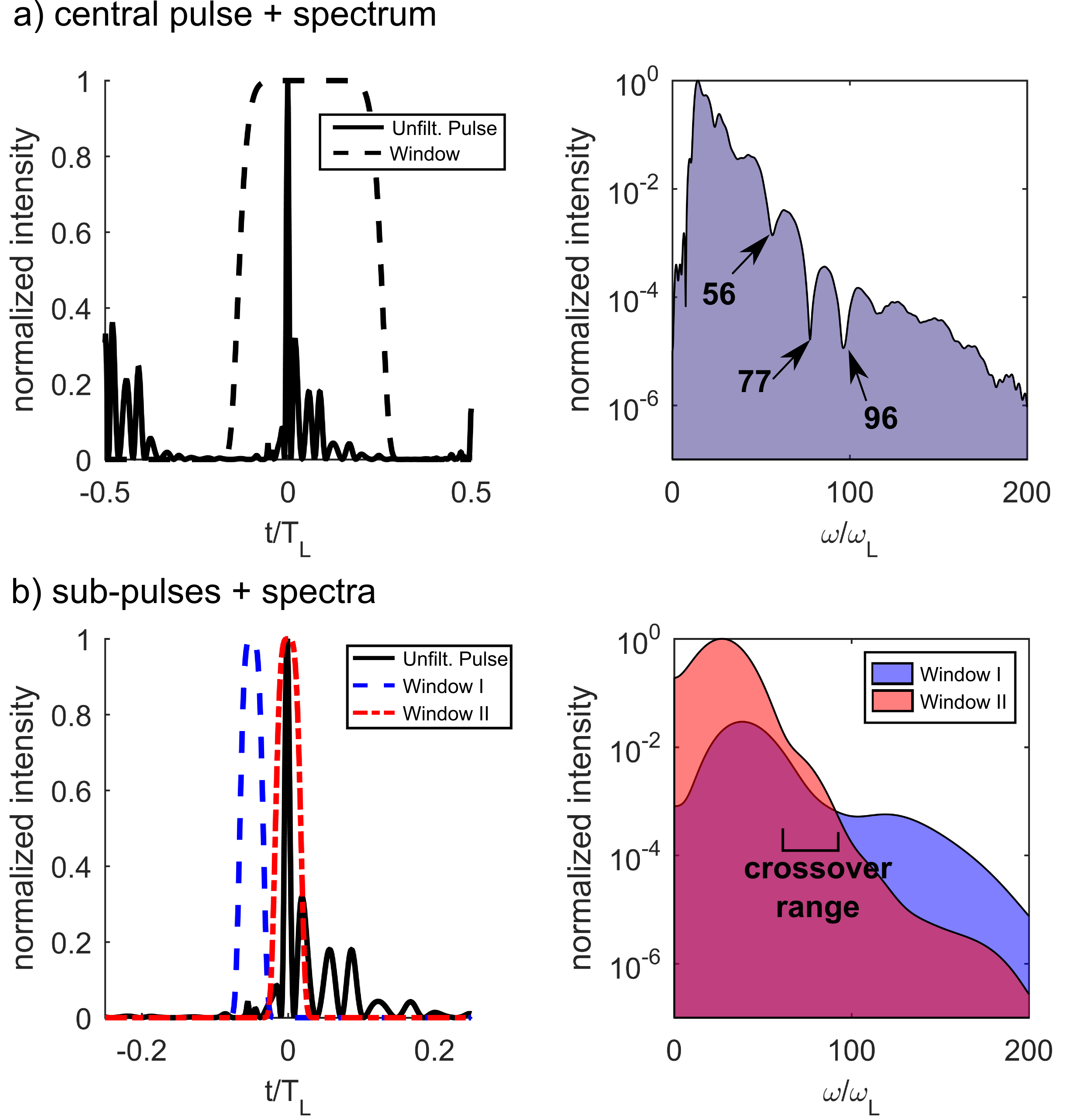}
\caption{Part a) shows that by selecting the peak pulse from the train, a broad spectrum of radiation is obtained.  However the modulations in the spectrum highlighted here, which are separated by $\Delta \omega \approx 20~\omega_L$, suggest the presence of a sub-pulse structure, with sources separated in time by $\Delta t\approx 130$ as.  Part b) shows that by selecting windows corresponding to the times of the sub-pulses (cf.\ Fig.~\ref{fig1}c)), the two constituent emission spectra can be obtained.  Here, it is further illustrated that one pulse dominates at high frequencies, whereas the other dominates at lower frequencies.  \label{fig3}}
\end{figure}

This substructure is also evident from the spectrum of the central pulse, shown in Fig.~\ref{fig3}a). The modulations indicated at $\omega/\omega_L \approx \{56, 77, 96\}$ result from interference between the two sub-pulses. Since these modulations are separated by $\Delta \omega/\omega_L \approx 20$, the temporal separation of the sub-pulses may be calculated as $\Delta t\approx130~\mathrm{as}$, in agreement with Fig.~\ref{fig1}c).  In part b), the two sub-pulses have been isolated by two supergaussian windows positioned over their peaks (the position of these peaks may be inferred from Fig.~\ref{fig1}c)).  The spectra of these two sub-pulses are shown, and it is clear that Window I, corresponding to emission from the primary electron bunch is dominant for $\omega/\omega_L >100$, whereas Window II, corresponding to emission from the secondary bunch is dominant for $\omega/\omega_L<60$.  For $60<\omega/\omega_L<100$ however, the two spectra are shown to be of comparable intensity, indicating that both electron bunches are contributing equally to the emission in this range.

\begin{figure}
\includegraphics[scale = 1]{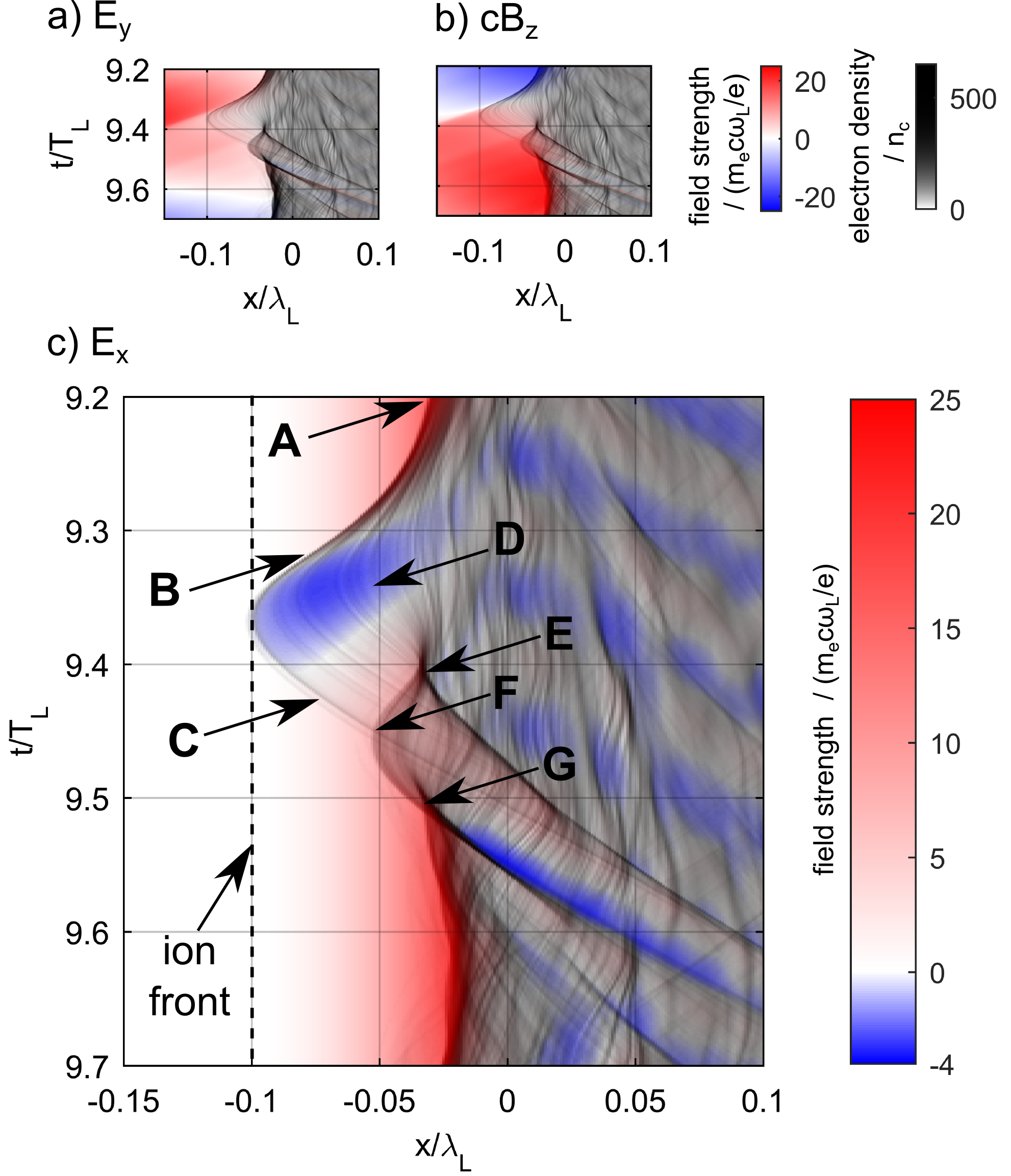}
\caption{The origin of the secondary emission bunch can be explained in terms of the electromagnetic fields driving its motion.  The electrons in the primary bunch at {\bf D} initiate a large amplitude electron oscillation, which ejects a secondary bunch of electrons at {\bf E} out towards the front surface of the plasma.  When this secondary bunch crosses the primary at {\bf F}, the electromagnetic fields of the driving laser pulse accelerate these electrons, triggering another burst of coherent synchrotron radiation at {\bf G}.  The formation of the primary bunch, points {\bf A} - {\bf C}, are described in the text.\label{fig4}}
\end{figure}

We now consider the dynamics of the two distinct bunches that contribute to transmitted CSE radiation.  The forces driving this process arise from the electromagnetic fields of the driving laser pulse ($E_y, B_z$), along with the longitudinal electric field component $E_x$, generated by the charged particles within the simulation.  In Fig.~\ref{fig4}, these fields are shown (red-blue) overlaid onto the electron density (gray-black), surrounding the point of emission.

Comparing Figs.~\ref{fig2alt} and \ref{fig4}, it is seen that the primary electron bunch is composed of those electrons at the front surface of the plasma and as such are directly influenced by the electromagnetic fields of the driving laser pulse. We now track the dynamics of this bunch, starting at $t/T_L=9.2$, the point marked {\bf A} on Fig.~\ref{fig4}c). At this point the $v_yB_z{\hat{\bf x}}$ component of the Lorentz force has driven the front-side electrons into the plasma, setting up a large electrostatic $E_x$ field.  This restoring force then accelerates the primary bunch to relativistic longitudinal velocities at point marked {\bf B}. Simultaneously, the $E_y$ field of the driving laser pulse provides a transverse acceleration which curves the trajectory of this electron bunch, resulting in the sole emission of the high frequency burst of synchrotron radiation in the reflected direction.  This primary bunch is then accelerated back towards the target by the $v_yB_z{\hat{\bf x}}$ force and attains a relativistic velocity in the forward direction at point {\bf C}.  This time, it is the $v_xB_z{\hat{\bf y}}$ component which provides a transverse acceleration, and results in a burst of synchrotron radiation in the transmitted direction.  

The origins of the secondary bunch begin at $t/T_L = 9.35$, at the point marked {\bf D}. As the primary bunch of electrons are accelerated away from the target, it generates a longitudinal electric field which pushes a fraction of its constituent electrons back into a target.  These electrons then bunch and eject a second set of electrons at point {\bf E} towards the front of the plasma.  As this secondary bunch of electrons cross the first, at {\bf F}, they become subject to the electromagnetic fields of the laser pulse which turn the bunch around, accelerating the electrons in the forward direction and triggering a secondary burst of synchrotron radiation at point {\bf G}.  

Note that immediately after point {\bf E}, bunches of electrons are accelerated to relativistic velocities in both forward and reflected directions yet do not emit significant levels of high frequency radiation.  The reason for this is that these electrons are shielded from the electromagnetic fields of the driving laser pulse, and as such do not experience significant transverse acceleration, which is a requirement for the emission of synchrotron radiation.  We also note that since these secondary bunches are accelerated to large velocities before being injected into the plasma, it is likely they make an important contribution to the total energy absorption of the target.

\begin{figure}
\includegraphics[scale = 0.8]{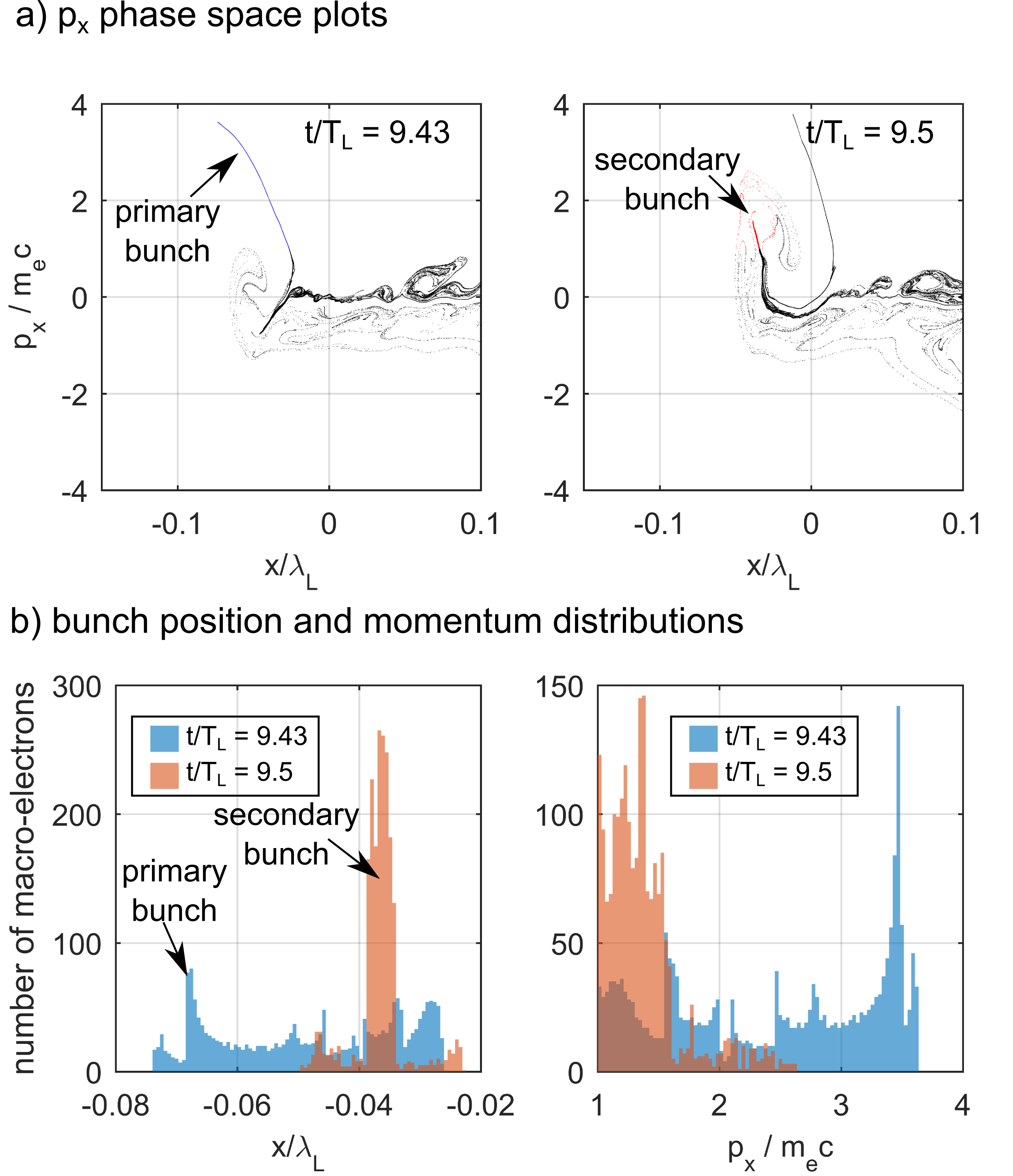}
\caption{Part a) shows the positions of the macro-electrons in $p_x$ phase space at the points in the simulation at which two consecutive sub-pulses are emitted. The electrons which constitute the primary and secondary bunches are highlighted, and are chosen as those electrons at the front surface of the target having $p_x/m_ec >1$. Part b) shows the position and momentum distributions of the these macro-electrons, which differ significantly for each of the two bunches.  The electrons in the primary bunch are shown to be more spread out in position, but typically posses a larger momentum than the secondary.  The electrons with the largest momenta in the primary bunch are positioned more towards the left hand side of the distribution.  The electrons forming the secondary bunch, though they typically have a lower forward momentum, are more densely positioned together over a $\approx$4 nm bunch. \label{fig5}}
\end{figure}

Having determined that two distinct electron bunches are responsible for the forward emissions, we now describe why the two pulses are dominant in different frequency ranges.  To do so, we look at the position and momentum distributions of the emitting electrons, shown in Fig.~\ref{fig5}. From the $p_x$ phase space plots shown in Fig.~\ref{fig5}a), it can be seen that the characteristics of these bunches are quite different at the point of emission.  To clarify, histograms showing the position and momentum distributions of the two bunches are shown in part b). Shown here are all the macro-electrons at the front surface of plasma during each emission process with a longitudinal momentum $p_x/m_ec\geq 1$.

Looking first at the $p_x$ histogram of the macro-electrons, it is clear that the majority of particles in the primary bunch have a forward momentum $p_x/m_ec\approx 3.75$, whereas for the second bunch the most common value is $p_x/m_ec\approx 1.5$.  From the electron density plots (Figs.~\ref{fig2alt} and \ref{fig4}) it is clear that the primary electron bunch has been exposed to the forward electromagnetic forces for longer than the secondary, therefore achieving higher velocities.  Furthermore, from Fig.~\ref{fig4}c), it may be seen that the restoring force of the ions is weaker in the region that the primary bunch is accelerated, and thus may attain higher velocities than the secondary.

The second main difference between the characteristics of the two emitting bunches is in the position distribution of the electrons during the emission, again displayed here in a histogram in Fig.~\ref{fig5}b).  Those electrons with $p_x/m_ec\geq 1$ in the primary bunch have a broad spatial distribution.  However, by comparison with Fig.~\ref{fig2alt} it is seen that the electrons which result in the majority of the primary emission are those towards the left of this distribution.  This is partly because of their higher energies (as shown in Fig.~\ref{fig5}a)), but also because they are closer to the front surface of the plasma and are more exposed to the electromagnetic fields of the laser pulse.  The primary bunch, having a broad spatial distribution, has a low peak density. In contrast, the majority of the electrons forming the secondary bunch are positioned closely together in a bunch of width $\approx 4$ nm.  This can be explained by looking at the longitudinal electric field in Fig.~\ref{fig4}.  Here it can be seen that the primary bunch disperses as its constituent electrons repel each other (i.e.\ they are accelerated forward by the $E_x$ field (blue)) leading to a reduction in density of this electron bunch.  The secondary bunch however experiences less of this dispersion and therefore maintains its density.

To summarize, these simulations suggest that the electrons in the primary bunch typically possess a larger longitudinal momentum $p_x$ than those in the secondary, however fewer electrons are contributing to the emission. These electron distributions therefore explain why the primary bunch dominates at the highest frequencies, whereas the secondary bunch dominates the lower.  In particular, electrons in the primary bunch, with their typically larger $p_x$ radiate to higher frequencies than the secondary bunch, and so are dominant in this region.  The electrons in the secondary bunch however, though the maximum frequency they radiate is lower, dominate the lower frequencies because there are more electrons in this bunch contributing coherently to the emission.


In conclusion, we have shown that coherent synchrotron emission from relativistic laser-nanofoil interactions can originate from two distinct electron bunches with a separate formation history. The first bunch is initiated by the laser fields, while the second is induced by the electrostatic field from the first. Emission from these two bunches have distinct spectral signatures with XUV emission of comparable strength at intermediate photon energies, resulting in a double-pulse structure on the attosecond timescale. Very short pulses ($\tau\approx70~\mathrm{as}$) are observed for optimally filtered pulses in either the low or high frequency domain of the emission spectrum.

\end{document}